\documentclass[amsmath,aps,showpacs,a4paper,10pt]{revtex4}
 \usepackage{epsf}
 \usepackage{graphicx}    % Include figure files
 \usepackage{enumerate}   % enumerate itemize

\begin{document}

 \newcommand{\be}[1]{\begin{equation}\label{#1}}
 \newcommand{\ee}{\end{equation}}
 \newcommand{\bea}{\begin{eqnarray}}
 \newcommand{\eea}{\end{eqnarray}}
 \def\disp{\displaystyle}

% Dear copy-editor,
% Please keep \dmunit and \cmn as they are. Otherwise, we will ask you
% to recover them in the proofreading. So, to save our and your time,
% please do NOT change them throughout this .tex file. Thanks.
%
 \newcommand{\dmunit}{ {\rm pc\hspace{0.24em} cm^{-3}} }
 \newcommand{\cmn}{,\,}

 \begin{titlepage}

 \begin{flushright}
 arXiv:2102.00579
 \end{flushright}

 \title{\Large \bf Effect of Redshift Distributions of Fast
 Radio Bursts on~Cosmological Constraints}

 \author{Da-Chun~Qiang\,}
 \email[\,email address:\ ]{875019424@qq.com}
 \affiliation{School of Physics,
 Beijing Institute of Technology, Beijing 100081, China}

 \author{Hao~Wei\,}
 \email[\,Corresponding author;\ email address:\ ]{haowei@bit.edu.cn}
 \affiliation{School of Physics,
 Beijing Institute of Technology, Beijing 100081, China}

 \begin{abstract}\vspace{1cm}
 \centerline{\bf ABSTRACT}\vspace{2mm}
 Nowadays, fast radio bursts (FRBs) have been a promising probe for
 astronomy and cosmology. However, it is not easy to identify the redshifts
 of FRBs to date. Thus, no sufficient actual FRBs with identified redshifts
 can be used to study cosmology currently. In the past years, one has to
 use the simulated FRBs with ``\,known\,'' redshifts instead. To simulate
 an FRB, one should randomly~assign a redshift to it from a given redshift
 distribution. But the actual redshift distribution of FRBs is still
 unknown so far. Therefore, many redshift distributions have been assumed
 in the literature. In the present work, we study the effect of various
 redshift distributions on cosmological constraints, while they are treated
 equally. We find that different redshift distributions lead to different
 cosmological constraining abilities from the simulated FRBs. This result
 emphasizes the importance to find the actual redshift distribution of
 FRBs, and reminds us of the possible bias in the FRB simulations due
 to the redshift distributions.
 \end{abstract}

% https://publishing.aip.org/publishing/pacs/pacs-2010-regular-edition
% https://publishing.aip.org/pacs_2010_alpha/
% https://publishing.aip.org/wp-content/uploads/2019/01/PACS_2010_Alpha.pdf
% https://ufn.ru/en/pacs/all/

 \maketitle

 \end{titlepage}

 \renewcommand{\baselinestretch}{1.0}

%============================= section 1 ===================================

\section{Introduction}\label{sec1}

Currently, fast radio bursts (FRBs) have become a thriving field in
 astronomy and cosmology~\cite{NAFRBs,Lorimer:2018rwi,Keane:2018jqo,
 Kulkarni:2018ola,Burke-Spolaor:2018xoa,Pen:2018ilo,Macquart:2018fhn,
 Caleb:2018ygr,Zhang:2020qgp,Xiao:2021omr,Petroff:2019tty,Cordes:2019cmq}.
 Since their first discovery~\cite{Lorimer:2007qn,Thornton:2013iua}, an
 extragalactic/cosmological origin is strongly suggested to FRBs, due
 to the large dispersion measure (DM) of observed FRBs well in excess
 of the Galactic value. To date, the redshifts of several FRBs have
 been identified by the precise localizations of their host
 galaxies~\cite{Tendulkar:2017vuq,Bannister:2019,Ravi:2019alc,
 Marcote:2020ljw,Heintz:2020,Macquart:2020lln,Bhandari:2020cde,
 Prochaska:2019}. For example, the redshift of the first known
 repeating FRB (namely FRB 121102) has been identified as
 $z=0.19273$~\cite{Tendulkar:2017vuq}. Currently, FRB 190523 has the
 largest identified redshift $z=0.66$~\cite{Ravi:2019alc}. The 12 FRBs
 with identified redshifts as of November 2020 were summarized
 in e.g.~\cite{Batten:2020pdh}. Clearly, they are all at cosmological
 distances. Therefore, it is justified and well motivated to
 study cosmology by using FRBs. We refer to e.g.~\cite{Deng:2013aga,
 Yang:2016zbm,Gao:2014iva,Zhou:2014yta,Yu:2017beg,Yang:2017bls,Wei:2018cgd,
 Li:2017mek,Jaroszynski:2018vgh,Madhavacheril:2019buy,Wang:2018ydd,
 Walters:2017afr,Qiang:2019zrs,Qiang:2020vta,Cai:2019cfw,Li:2019klc,
 Wei:2019uhh} for some interesting works on the FRB cosmology.

As is well known, one of the key observational quantities of FRBs is the
 dispersion measure DM. The radio signals of different frequencies from FRB
 reach earth at different times, due to the cold plasma along the path.
 According to e.g.~\cite{Rybicki:1979}, in the rest frame, an
 electromagnetic signal propagates through an ionized medium (plasma) with
 a velocity less than the speed of light in vacuum~$c$, and hence this
 signal of frequency $\nu\gg\nu_p$ is delayed relative to a signal in vacuum
 by a time proportional to $\nu^{-2}$ and the column density of the free
 electrons, where $\nu_p$ is the plasma frequency. In practice, it is
 convenient to measure the time delay $\Delta t$ in the observer frame
 between two signals of frequencies $\nu_1$ and $\nu_2$. Taking the redshift
 effect into account, this time delay is given by~\cite{Deng:2013aga,
 Yang:2016zbm,Ioka:2003fr,Inoue:2003ga,Qiang:2019zrs,Qiang:2020vta}
 \be{eq1}
 \Delta t=\frac{e^2}{2\pi m_{e\,}c}\left(
 \frac{1}{\nu_{1}^2}-\frac{1}{\nu_{2}^2}\right)
 \int\frac{n_{e\cmn z}}{1+z}\,dl\equiv\frac{e^2}{2\pi m_{e\,}c}\left(
 \frac{1}{\nu_{1}^2}-\frac{1}{\nu_{2}^2}\right){\rm DM}\,,
 \ee
 where $n_{e\cmn z}$ is the number density of free electrons in the medium
 (given in units of $\rm cm^{-3}$) at redshift $z$, $m_e$ and $e$ are the
 mass and charge of electron, respectively. Using Eq.~(\ref{eq1}), one can
 get the column density of the free electrons ${\rm DM}\equiv
 \int n_{e\cmn z}/(1+z)\,dl$ by measuring the time delay $\Delta t$ between
 two signals of frequencies $\nu_1$ and $\nu_2$. It is worth noting that
 the distance $dl$ along the path in DM records the expansion history of
 the universe. Therefore, DM plays a key role in the FRB cosmology.

Clearly, the observed DM of FRB can be separated into~\cite{Deng:2013aga,
 Yang:2016zbm,Qiang:2019zrs,Qiang:2020vta,Gao:2014iva,Zhou:2014yta,
 Yang:2017bls,Li:2019klc,Wei:2019uhh}
 \be{eq2}
 \rm DM_{obs}=DM_{MW}+DM_{IGM}+DM_{HG}\,,
 \ee
 where $\rm DM_{MW}$, $\rm DM_{IGM}$, and $\rm DM_{HG}$ are the
 contributions from the Milky Way, the intergalactic medium (IGM), and the
 host galaxy (HG, including interstellar medium of HG and the near-source
 plasma), respectively. Since thousands of pulsars in the Milky Way and
 the Small/Large Magellanic Clouds were observed, one can reliably infer
 the density distribution of the free electrons in or nearby the Milky Way
 from the observed DMs of these pulsars. So, for a well-localized FRB, the
 corresponding $\rm DM_{MW}$ can be estimated with reasonable certainty by
 using the well-known tools NE2001~\cite{Cordes:2002wz,Cordes:2003ik} or
 YMW16~\cite{YMW16}. Thus, subtracting the ``\,known\,'' $\rm DM_{MW}$
 from $\rm DM_{obs}$ in Eq.~(\ref{eq2}), it is convenient to introduce the
 extragalactic DM of an FRB as the observed quantity~\cite{Yang:2016zbm,
 Yang:2017bls,Qiang:2019zrs,Qiang:2020vta,Li:2019klc},
 \be{eq3}
 \rm DM_E\equiv DM_{IGM}+DM_{HG}\,.
 \ee
 The main contribution to DM of FRB comes from IGM. In fact, $\rm DM_{IGM}$
 carries the key information about IGM and the cosmic expansion history. In
 principle, one can constrain cosmological models by using the
 observed $\rm DM_E$ of a large number of FRBs with identified redshifts.

Unfortunately, it is not so easy to identify the redshifts of FRBs to date.
 Since the first discovery of FRB~\cite{Lorimer:2007qn,Thornton:2013iua},
 the redshifts have been identified only for 12 extragalactic FRBs, as is
 summarized in e.g.~\cite{Batten:2020pdh}. Therefore, no sufficient actual
 FRBs with identified redshifts can be used to study cosmology currently.
 In the past years, one has to use the simulated FRBs with ``\,known\,''
 redshifts instead. The devil is in the details. To simulate an FRB, one
 should randomly assign a redshift to it from a given redshift distribution.
 However, the actual redshift distribution of FRBs is still unknown to
 date. Therefore, various redshift distributions have been assumed in
 the literature, while some of them are motivated by the star formation
 history/rate~\cite{Munoz:2016tmg,Bhattacharya:2020rtf,Zhang:2020ass,
 James:2021oep} or compact binary mergers~\cite{Zhang:2020ass} and so on,
 some of them are borrowed from other objects such as gramma-ray bursts
 (GRBs) \cite{Yang:2016zbm,Gao:2014iva,Zhou:2014yta,Walters:2017afr,
 Wei:2019uhh,Qiang:2019zrs}, some of them come from the observed FRBs (such
 as Burr and Burr12 proposed in this work), and some of them are not well
 motivated at all (e.g.~Uniform). In the present work, we are interested to
 see whether or not various redshift distributions used to simulate FRBs
 can affect cosmological constraints considerably, and we do not care
 whether these redshift distributions are well motivated or where they
 come from. Our goal is just to see how they affect the cosmological
 constraints, while they are treated equally in this work, no matter
 whether they are the intrinsic ones or the observed ones.

The rest of this paper is organized as follows. In Sec.~\ref{sec2}, we
 introduce various redshift distributions for FRBs considered extensively in
 the literature. In addition, we also propose two new redshift distributions
 inferred from the actual FRBs data to date, which are fairly different from
 the existing ones in the literature. In Sec.~\ref{sec3}, we briefly
 describe the key points to simulate FRBs. In Sec.~\ref{sec4}, we constrain
 various cosmological models by using these simulated FRBs, and try to see
 the effect of redshift distributions on cosmological constraints. In
 Sec.~\ref{sec5}, some brief concluding remarks are given.

%============================= section 2 ===================================

\section{Various redshift distributions for FRBs}\label{sec2}

%============================= section 2a ===================================

\subsection{New redshift distributions}\label{sec2a}

\begin{center}
 \begin{figure}[tb]
 \centering
 \vspace{-8mm}  % used here just for a comfortable typesetting
 \includegraphics[width=0.85\textwidth]{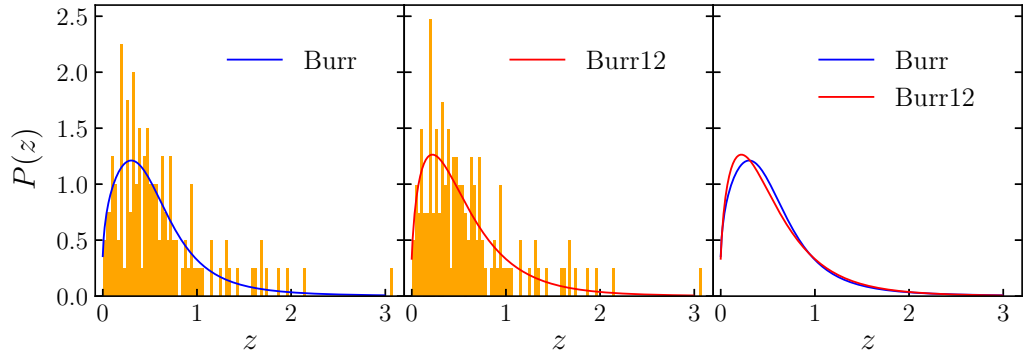}
 \caption{\label{fig1} Left panel: The best Burr distribution
 (blue solid line) as a function of redshift $z$ versus the normalized
 histogram of 129 actual FRBs with inferred redshifts plus a Galactic
 FRB 200428 at $z=0$. Middle panel: The best Burr Type XII distribution
 (red solid line) as a function of redshift $z$ versus the normalized
 histogram of 118 actual FRBs with inferred redshifts and 12 actual FRBs
 with observed redshifts plus a Galactic FRB 200428 at $z=0$. Right
 panel: Comparison of the best Burr and Burr Type XII distributions.
 See Sec.~\ref{sec2a} for details.}
 \end{figure}
 \end{center}

In the literature, various redshift distributions for FRBs have been
 extensively considered. To our best knowledge, (almost) all of them are
 not inferred from the actual FRBs data. So, let us try it at first. Note
 that an online catalogue of the observed FRBs can be found in
 FRBCAT~\cite{Petroff:2016tcr}, which summarizes almost all observational
 aspects concerning the published FRBs. As of January 2021, FRBCAT catalogue
 contains 129 observed FRBs. Of course, most of them have no identified
 redshifts. However, one can roughly infer the redshift from the observed DM
 of FRB, following the methodology described in e.g. Sec.~2.2
 of~\cite{Hashimoto:2019aqu}. Since this is an inferred redshift, we do
 not require a high precision, and hence we can slightly simplify the
 methodology of~\cite{Hashimoto:2019aqu}. For an observed FRB,
 its $\rm DM_{obs}$ can be separated into three components as
 in Eq.~(\ref{eq2}). One can directly read its $\rm DM_{MW}$
 from FRBCAT~\cite{Petroff:2016tcr}, which is estimated by
 using NE2001~\cite{Cordes:2002wz,Cordes:2003ik} or YMW16~\cite{YMW16}. On
 the other hand, one can assume ${\rm DM_{HG}}=50/(1+z)\;\dmunit$ following
 e.g.~\cite{Hashimoto:2019aqu,Shannon:2018}. The mean $\rm DM_{IGM}$ can
 be estimated by ${\rm DM_{IGM}}=3cH_0\Omega_b/(8\pi G m_p)\int_0^z
 (f_{\rm IGM}(\tilde{z}) f_e(\tilde{z}) (1+\tilde{z})/E(\tilde{z}))\,
 d\tilde{z}$ (see e.g.~\cite{Deng:2013aga,Yang:2016zbm,Qiang:2019zrs,
 Qiang:2020vta,Li:2019klc,Wei:2019uhh,Hashimoto:2019aqu}), in which one can
 assume the simplest flat $\Lambda$CDM cosmology to estimate $E(z)=
 (\Omega_m(1+z)^3+(1-\Omega_m))^{1/2}$ and adopt the values of $\Omega_m$,
 $\Omega_b$, $H_0$ from Planck 2018 results~\cite{Aghanim:2018eyx},
 while $f_{\rm IGM}=0.83$ and $f_e=7/8$ as in e.g.~\cite{Deng:2013aga,
 Yang:2016zbm,Yang:2017bls,Gao:2014iva,Qiang:2019zrs}. So, the right hand
 side of Eq.~(\ref{eq2}) becomes an explicit function of redshift $z$. For
 an observed FRB, one can infer its redshift $z$ by numerically solving
 Eq.~(\ref{eq2}) with the observational value of $\rm DM_{obs}$. Of course,
 we stress that it is just a roughly inferred redshift only for reference.
 Following this methodology, now we have 129 actual FRBs with inferred
 redshifts. To get a reasonable redshift distribution, we need an anchor.
 Very recently, FRB 200428 in our Milky Way was observed (see
 e.g.~\cite{Andersen:2020hvz,Bochenek:2020zxn,Lin:2020mpw,Li:2020qak}).
 So, we also take this FRB at $z=0$ into account. We fit these 130 actual
 FRBs with the {\bf fitter} Python package~\cite{fitter}, which can
 find the most probable distribution(s) for a given data sample by using
 80 distributions in SciPy~\cite{scipy}. Finally, we obtain the best
 redshift distribution for these 130 actual FRBs, namely Burr
 distribution~\cite{burr} (note that Burr Type III distribution is called
 Burr distribution for short in SciPy). The standardized
 Burr distribution is given by~\cite{burr}
 \be{eq4}
 f_{\rm\, Burr}(x\cmn b\cmn k)
 =\frac{bk\, x^{-b-1}}{\left(1+x^{-b}\right)^{k+1}}\,,
 \ee
 where $x\geq 0$, $b>0$ and $k>0$. One can shift and/or scale this
 distribution by using the shift and scale parameters ($\ell$ and $s$),
 namely~\cite{burr}
 \be{eq5}
 P_{\rm\, Burr}(z\cmn b\cmn k\cmn \ell\cmn s)=
 f_{\rm\, Burr}((z-\ell)/s\cmn b\cmn k)\,/s\,.
 \ee
 The best parameters for the 130 actual FRBs mentioned above
 are $b=2.8733$, $k=0.4568$, $\ell=-0.0043$ and $s=0.7357$. We present this
 best Burr distribution in the left panel of Fig.~\ref{fig1}.

As mentioned above, there are 12 extragalactic FRBs with identified
 redshifts to date, as is summarized in e.g. Table~2 of
 \cite{Batten:2020pdh}. Eleven of them are also compiled in the above
 129 FRBs catalogue, while FRB 200430 is not. We replace the inferred
 redshifts of these 11 FRBs by the actually identified ones, and also
 take FRB 200430 into account. Similarly, we fit these 131 FRBs with
 the {\bf fitter} Python package~\cite{fitter}, and then obtain the
 best redshift distribution, namely Burr Type XII distribution (Burr12)
 \cite{burr12}. The standardized Burr Type XII distribution is given
 by~\cite{burr12}
 \be{eq6}
 f_{\rm\, Burr12}(x\cmn b\cmn k)=
 \frac{bk\, x^{b-1}}{\left(1+x^b\right)^{k+1}}\,,
 \ee
 where $x\geq 0$, $b>0$ and $k>0$. One can shift and/or scale this
 distribution by using the shift and scale parameters ($\ell$ and $s$),
 namely~\cite{burr12}

%============================= Fig. 1 =================================

%======================================================================

 \vspace{-14.5mm}  % used here just for a comfortable typesetting

 \be{eq7}
 P_{\rm\, Burr12}(z\cmn b\cmn k\cmn \ell\cmn s)=
 f_{\rm\, Burr12}((z-\ell)/s\cmn b\cmn k)\,/s\,.
 \ee
 The best parameters for the 131 actual FRBs mentioned above
 are $b=1.4653$, $k=3.9060$, $\ell=-0.0064$ and $s=1.3963$. We present this
 best Burr Type XII distribution in the middle panel of Fig.~\ref{fig1}.

We compare the best Burr and Burr12 distributions obtained above in the
 right panel of Fig.~\ref{fig1}. It is easy to see that they are fairly
 close in fact. We stress that these two new redshift distributions Burr and
 Burr12 are not the actual one of FRBs, since the inferred redshifts are
 rough, while the 129 observed FRBs from FRBCAT are collected from many
 different telescopes with different sensitivities, band widths, central
 frequencies, fields of view, and operation times. Thus, many selection
 effects exist in these 129 observed FRBs (we thank the referee for pointing
 out this issue). On the other hand, these two new redshift distributions
 Burr and Burr12 are the observed ones, which are different from the
 intrinsic ones. One should be aware of this. However, as mentioned in
 Sec.~\ref{sec1}, our goal is just to see how redshift distributions affect
 the cosmological constraints, and hence they are treated equally in this
 work, no matter whether they are the intrinsic ones or the observed ones,
 and we do not care whether these redshift distributions are well motivated
 or where they come from. So, Burr and Burr12 do not have any special
 position. Let us be very clear that they are just trivial two in all
 the 9 redshift distributions considered equally in this work.

%============================= Fig. 2 =================================

 \begin{center}
 \begin{figure}[tb]
 \centering
 \vspace{-7mm}  % used here just for a comfortable typesetting
 \includegraphics[width=0.85\textwidth]{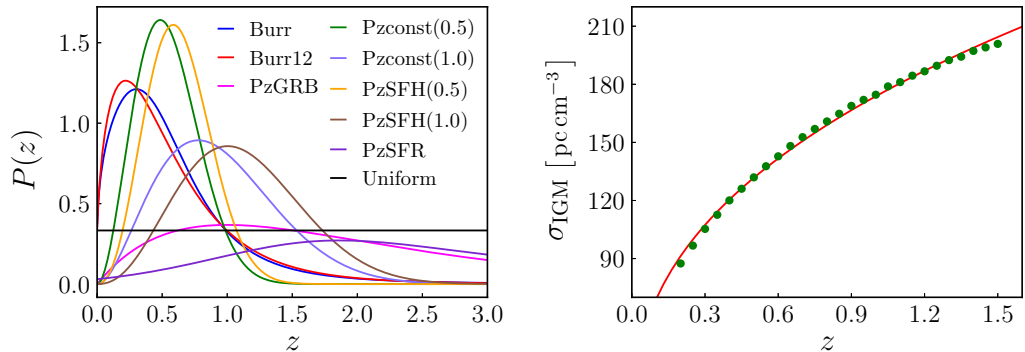}
 \caption{\label{fig2} Left panel: The 9 normalized redshift distributions
 for FRBs. See Sec.~\ref{sec2} for details. Right panel: $\sigma_{\rm IGM}$
 versus redshift $z$. The 27 green dots are reproduced from the bottom
 panel of Fig.~1 of~\cite{McQuinn:2013tmc}. The red solid line
 is plotted according to Eq.~(\ref{eq16}). See Sec.~\ref{sec3} for details.}
 \end{figure}
 \end{center}

%======================================================================

 \vspace{-12.9mm}  % used here just for a comfortable typesetting

%============================= section 2b ===================================

\subsection{Existing redshift distributions}\label{sec2b}

In fact, many existing redshift distributions for FRBs have been extensively
 considered in the literature. Note that they do not come from the actual
 FRBs. Since the actual redshift distribution of FRBs is still unknown to
 date, one might borrow the ones of other objects. For example,
 in e.g.~\cite{Yang:2016zbm,Gao:2014iva,Zhou:2014yta,Walters:2017afr,
 Wei:2019uhh,Qiang:2019zrs}, one can argue that FRBs are similar/related
 to gamma-ray bursts (GRBs), and hence assume that the redshift distribution
 of FRBs takes the one of GRBs~\cite{Shao:2011xt} (termed ``\,PzGRB\,''),
 namely
 \be{eq8}
 P_{\rm GRB}(z)\propto z\, e^{-z}\,,
 \ee
 which is a special case of Erlang distribution~\cite{erlang}.
 PzGRB was used extensively in the literature.

In e.g.~\cite{Munoz:2016tmg}, two redshift distributions for FRBs were
 proposed. The first one (termed ``\,Pzconst\,'') assumes that FRBs have
 a constant comoving number density, and the corresponding redshift
 distribution function is given by~\cite{Munoz:2016tmg}

%============================= Fig. 3 =================================

 \begin{center}
 \begin{figure}[tb]
 \centering
 \vspace{-10mm} \hspace{-6mm} % used here just for a comfortable typesetting
 \includegraphics[width=0.82\textwidth]{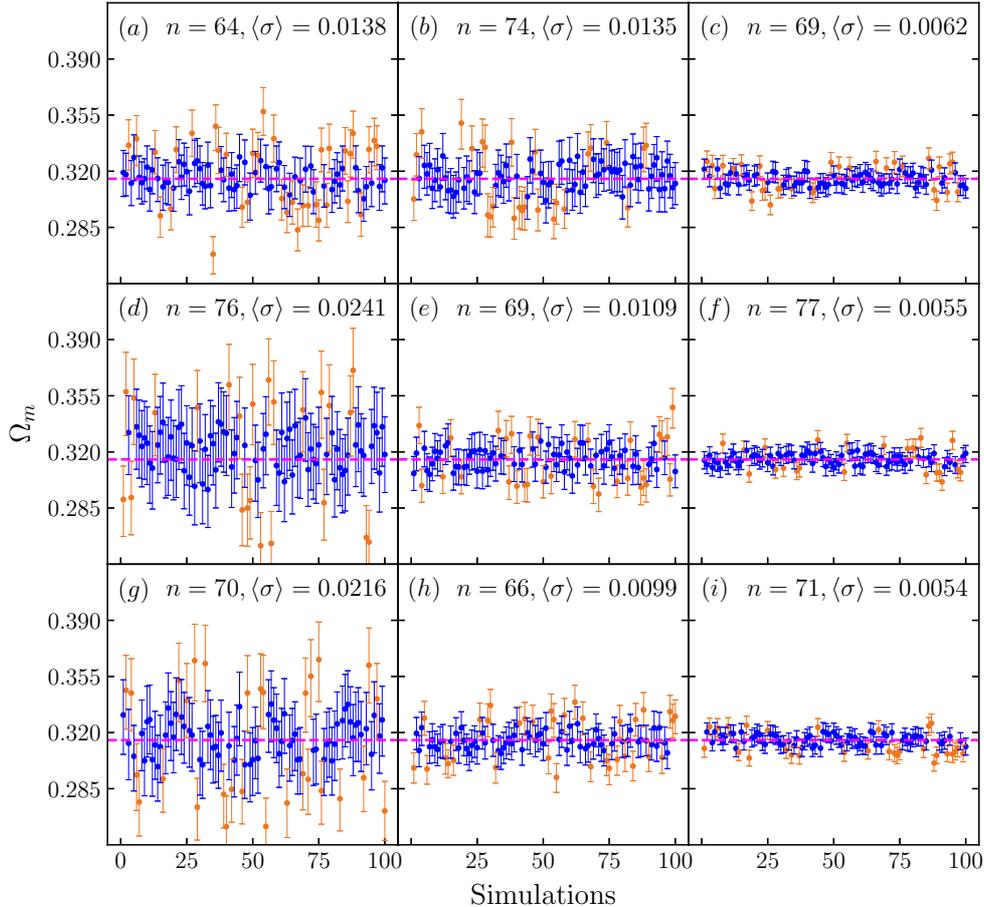}
 \caption{\label{fig3} Panels $(a)$ to $(i)$ correspond to the redshift
 distributions Burr, Burr12, PzGRB, Pzconst(0.5), Pzconst(1.0),
 PzSFR, PzSFH(0.5), PzSFH(1.0), and Uniform, respectively. In each panel,
 the marginalized $1\sigma$ constraints on the cosmological parameter
 $\Omega_m$ of the flat $\Lambda$CDM model for 100 simulations
 are presented. In each simulation, $N_{\rm FRB}=1000$ FRBs are generated
 by using the flat $\Lambda$CDM model with the preset parameter
 $\Omega_m=0.3153$ (indicated by the magenta dashed lines).
 The blue means with error bars (the chocolate means with error bars)
 indicate that the preset $\Omega_m=0.3153$ is consistent (inconsistent)
 with the simulated FRBs within $1\sigma$ region, respectively. $n$ and
 $100-n$ are the numbers of blue and chocolate means with error bars,
 respectively. $\langle\sigma\rangle$ is the mean of the uncertainties of
 100 constraints on the cosmological parameter $\Omega_m$. See
 Sec.~\ref{sec4} for details.}
 \end{figure}
 \end{center}

%======================================================================

%============================= Fig. 4 =================================

 \begin{center}
 \begin{figure}[tb]
 \centering
 \vspace{-10mm} \hspace{-6mm} % used here just for a comfortable typesetting
 \includegraphics[width=0.82\textwidth]{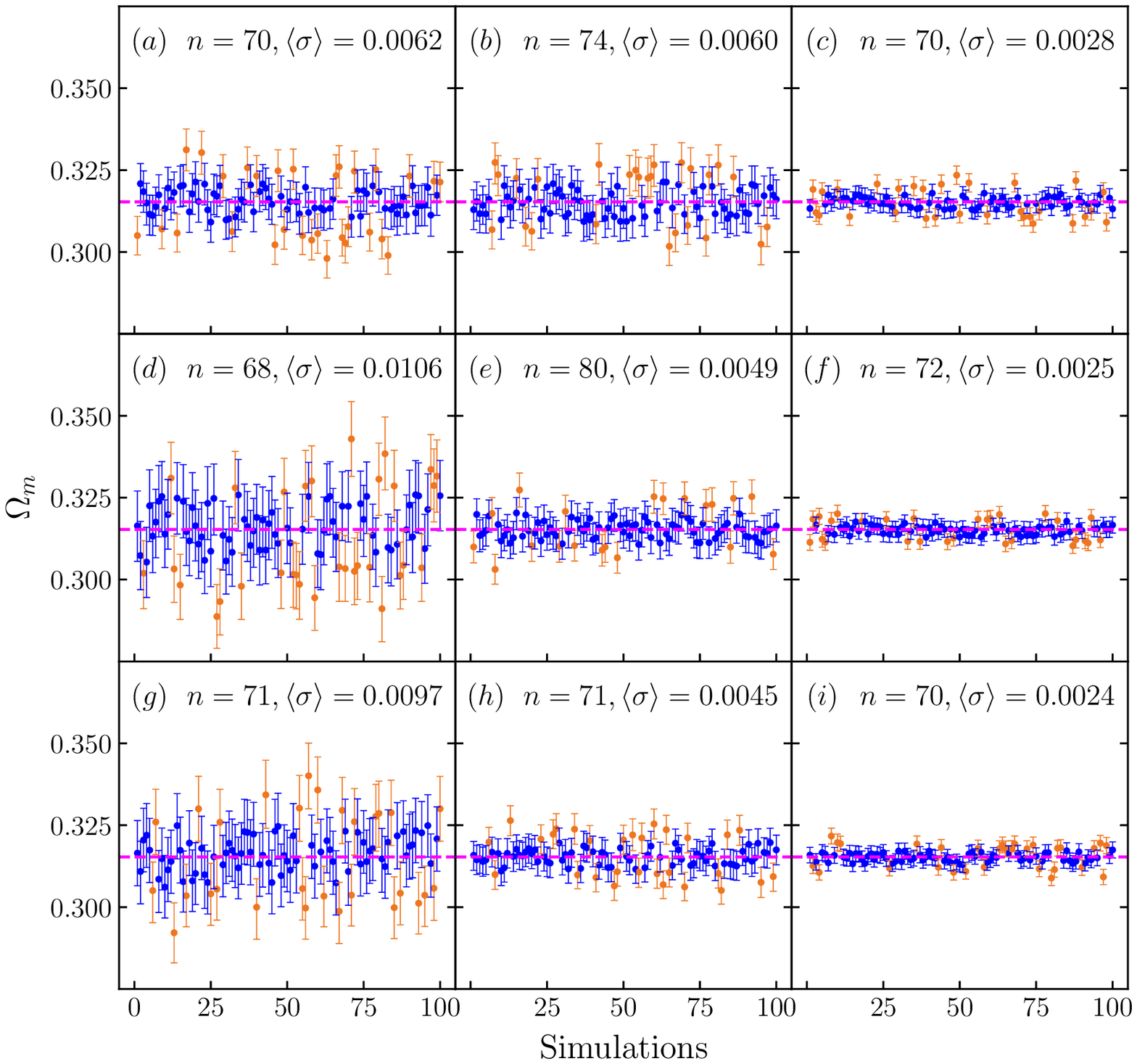}
 \caption{\label{fig4} The same as in Fig.~\ref{fig3},
 but $N_{\rm FRB}=5000$ in each simulation.}
 \end{figure}
 \end{center}

%======================================================================

 \vspace{-23mm}  % used here just for a comfortable typesetting for Figs. 3 and 4

 \be{eq9}
 P_{\rm const}(z)\propto\frac{d_C^2(z)}{\left(1+z\right)H(z)}
 \,\exp\left(-\frac{d_L^{\,2}(z)}{2\,d_L^{\,2}(z_{\rm cut})}\right)\,,
 \ee
 where $H(z)$ is the Hubble parameter, $d_C(z)=d_L(z)/(1+z)=c\int_0^z
 d\tilde{z}/H(\tilde{z})$ is the comoving distance, $d_L(z)$ is the
 luminosity distance. Gaussian cutoff at $z_{\rm cut}$ is introduced to
 represent an instrumental signal-to-noise threshold. The second one
 (termed ``\,PzSFH\,'') assumes that FRBs follow the star-formation
 history (SFH)~\cite{Caleb:2015uuk}, whose density is given by
 \be{eq10}
 \dot{\rho}_\ast(z)=\frac{\left(b_1+b_2 z\right)h}{\,1+
 \left(z/b_3\right)^{b_4\,}}\,,
 \ee
 with $b_1=0.0170$, $b_2=0.13$, $b_3=3.3$, $b_4=5.3$ and
 $h=0.7$~\cite{Cole:2000ea,Hopkins:2006bw,Munoz:2016tmg}. In this case, the
 SFH-based redshift distribution function reads~\cite{Munoz:2016tmg}
 \be{eq11}
 P_{\rm SFH}(z)\propto
 \frac{\dot{\rho}_\ast(z)\, d_C^2(z)}{\,\left(1+z\right)H(z)}\,
 \exp\left(-\frac{d_L^{\,2}(z)}{2\,d_L^{\,2}(z_{\rm cut})}\right)\,.
 \ee
 In the literature, the cutoff $z_{\rm cut}$ has been set to various
 values. In~\cite{Munoz:2016tmg} and e.g.~\cite{Qiang:2020vta,Wang:2018ydd},
 $z_{\rm cut}=0.5$ was adopted. On the other hand, $z_{\rm cut}=1.0$ was
 considered in e.g.~\cite{Li:2017mek,Li:2019klc}. We call the corresponding
 redshift distributions ``\,Pzconst(0.5)\,'', ``\,Pzconst(1.0)\,'',
 ``\,PzSFH(0.5)\,'', ``\,PzSFH(1.0)\,'', respectively.

Another type of redshift distribution for FRBs was proposed in
 e.g.~\cite{Bhattacharya:2020rtf}. One might argue that the distribution
 of FRBs closely trace the cosmic star-formation rate (SFR) for young
 stellar FRB progenitors. In e.g.~\cite{Madau:2014bja}, the cosmic SFR
 function is given by
 \be{eq12}
 \psi(z)=\frac{0.015 \left(1+z\right)^{2.7}}
 {\, 1+\left((1+z)/2.9\right)^{5.6}\,}\; M_\odot\;{\rm yr^{-1}\,Mpc^{-3}}\,.
 \ee
 The appropriately weighted redshift distribution is obtained
 by considering the quantity~\cite{Bhattacharya:2020rtf}
 \be{eq13}
 \zeta_{\rm\; SFR}=\frac{\int_0^z \psi(\tilde{z})\, d\tilde{z}}
 {\int_0^{z_{\rm max}} \psi(\tilde{z})\, d\tilde{z}}\,,
 \ee
 and drawing it as a uniform random number between $0$ and $1$. Since the
 right hand side of Eq.~(\ref{eq13}) is an explicit function of redshift
 $z$, for any uniform random number $0\leq\zeta_{\rm\; SFR}\leq 1$, one can
 obtain the corresponding redshift $z$ by numerically solving
 Eq.~(\ref{eq13}). So, the SFR-based redshift distribution
 (termed ``\,PzSFR\,'') can be generated for
 FRBs. In principle, $z_{\rm max}$ can be set to any value, and then PzSFR
 generates random redshifts in the range of $0\leq z\leq z_{\rm max}$.

Naively, since there is no guideline for the redshift distribution of FRBs
 to date, it is also reasonable to just consider a uniform distribution.
 One can uniformly assign a random redshift $z$ from 0 to $z_{\rm max}$.
 In the present work, we also take this uniform redshift distribution into
 account.

In the left panel of Fig.~\ref{fig2}, we summarize the 9
 redshift distributions for FRBs, which are all normalized. Notice that
 the distributions Pzconst and PzSFH are plotted just for demonstration
 by assuming the simplest flat $\Lambda$CDM cosmology with $\Omega_m=0.3153$
 taken from Planck 2018 results~\cite{Aghanim:2018eyx}. We stress that
 there are other types of redshift distributions for FRBs in the literature.
 We do not try to consider all redshift distributions for FRBs in a limited
 work. With these 9 redshift distributions, we simulate FRBs and then
 try to see the effect of redshift distributions for FRBs on cosmological
 constraints.

%============================= section 3 ===================================

\section{Simulating FRBs}\label{sec3}

Here, we briefly describe the key points to simulate FRBs. As
 mentioned in Sec.~\ref{sec1}, we consider the extragalactic DM defined in
 Eq.~(\ref{eq3}) as the observed quantity. The main contribution comes from
 IGM. As is shown in e.g.~\cite{Deng:2013aga,Yang:2016zbm,Qiang:2019zrs,
 Qiang:2020vta,Li:2019klc,Wei:2019uhh}, the mean of $\rm DM_{IGM}$ is
 given by
 \be{eq14}
 \langle{\rm DM_{IGM}}\rangle=\frac{3cH_0\Omega_{b}}{8\pi G m_p}
 \int_0^z\frac{f_{\rm IGM}(\tilde{z})\,f_e(\tilde{z})\left(1+
 \tilde{z}\right)d\tilde{z}}{E(\tilde{z})}\,,
 \ee
 where $\Omega_b$ is the present fractional density of baryons, $m_p$ is
 the mass of proton, $H_0$ is the Hubble constant, $E\equiv H/H_0$ is the
 dimensionless Hubble parameter. $f_{\rm IGM}$ is the fraction of baryon
 mass in IGM, which is a function of redshift $z$ in principle
 \cite{Li:2019klc,Wei:2019uhh,Qiang:2020vta}. Following
 e.g.~\cite{Yang:2016zbm,Yang:2017bls,Gao:2014iva,Qiang:2019zrs}, here we
 adopt a constant $f_{\rm IGM}=0.83$ (see e.g.~\cite{Fukugita:1997bi,
 Shull:2011aa} and \cite{Deng:2013aga}). The ionized electron number
 fraction per baryon is
 \be{eq15}
 f_e(z)\equiv Y_{\rm H}\,\chi_{e\cmn\rm H}(z)+
 \frac{1}{2}\,Y_{\rm He}\,\chi_{e\cmn\rm He}(z)\,,
 \ee
 in which hydrogen (H) and helium (He) mass fractions are
 $Y_{\rm H}=(3/4)\,y_1$ and $Y_{\rm He}=(1/4)\,y_2$, where $y_1\sim 1$
 and $y_2\simeq 4-3y_1\sim 1$ are the hydrogen and helium mass fractions
 normalized to the typical values $3/4$ and $1/4$, respectively. In
 principle, the ionization fractions $\chi_{e\cmn\rm H}(z)$ and
 $\chi_{e\cmn\rm He}(z)$ are both functions of redshift~$z$. It is expected
 that intergalactic hydrogen and helium are fully ionized at redshifts
 $z\lesssim 6$ and $z\lesssim 3$ \cite{Meiksin:2007rz,Becker:2010cu}
 (see also e.g.~\cite{Shull:2010ku,Beniamini:2020ane}), respectively. Thus,
 for FRBs at redshifts $z\leq 3$, they are both fully ionized,
 namely $\chi_{e\cmn\rm H}(z)=\chi_{e\cmn\rm He}(z)=1$. So,
 $f_e(z)\simeq 7/8$ for $z\leq 3$.

Note that $\rm DM_{IGM}$ will deviate from
 the mean $\langle{\rm DM_{IGM}}\rangle$ if the plasma density
 fluctuations are taken into account~\cite{McQuinn:2013tmc} (see also
 e.g.~\cite{Ioka:2003fr,Jaroszynski:2018vgh}). The uncertainty
 $\sigma_{\rm IGM}$ was studied in e.g.~\cite{McQuinn:2013tmc}, where three
 models for halo gas profile of the ionized baryons were used. Following
 e.g.~\cite{Qiang:2020vta}, we consider the simplest one, namely the top hat
 model, and the corresponding $\sigma_{\rm IGM}$ was given by the green dots
 in the bottom panel of Fig.~1 of~\cite{McQuinn:2013tmc}. It is easy to fit
 these 27 green dots with a simple power-law function~\cite{Qiang:2020vta}
 \be{eq16}
 \sigma_{\rm IGM}(z)=173.8\, z^{0.4}\ \dmunit\,.
 \ee
 In the right panel of Fig.~\ref{fig2}, we reproduce these 27 green
 dots from~\cite{McQuinn:2013tmc}, and also plot the power-law
 $\sigma_{\rm IGM}(z)$ given by Eq.~(\ref{eq16}). Obviously,
 they coincide with each other fairly well.

The contribution from the host galaxy of FRB, i.e. $\rm DM_{HG}$, is
 poorly known. The observed $\rm DM_{HG}$ for an FRB at
 redshift $z$ is given by (e.g.~\cite{Yang:2016zbm,Gao:2014iva,Yang:2017bls,
 Zhou:2014yta,Qiang:2019zrs,Qiang:2020vta,Li:2019klc})

%============================= Fig. 5 =================================

 \begin{center}
 \begin{figure}[tb]
 \centering
 \vspace{-8mm} \hspace{-6mm} % used here just for a comfortable typesetting
 \includegraphics[width=0.82\textwidth]{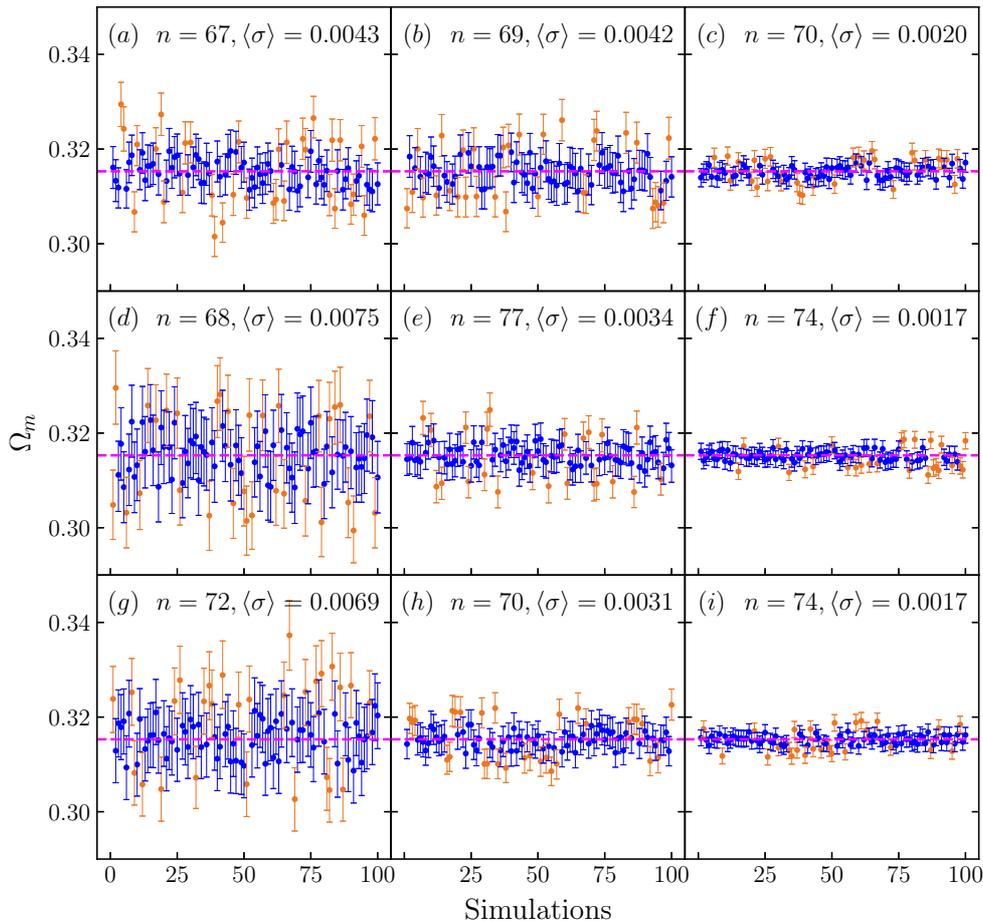}
 \caption{\label{fig5} The same as in Fig.~\ref{fig3},
 but $N_{\rm FRB}=10000$ in each simulation.}
 \end{figure}
 \end{center}

%======================================================================

%============================= Fig. 6 =================================

 \begin{center}
 \begin{figure}[tb]
 \centering
 \vspace{-8mm} \hspace{-6mm} % used here just for a comfortable typesetting
 \includegraphics[width=0.82\textwidth]{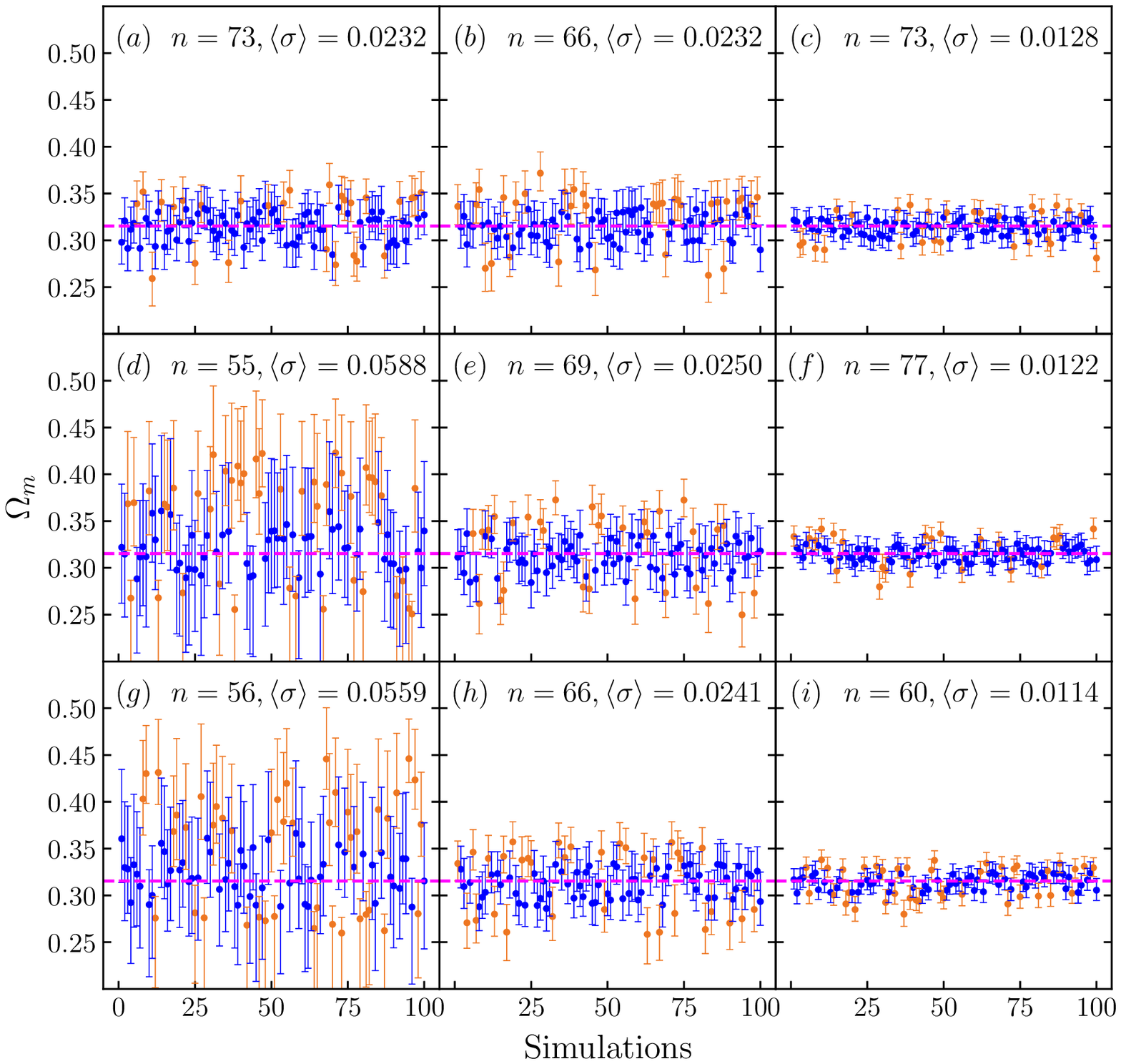}
 \caption{\label{fig6} The same as in Fig.~\ref{fig3}, but in
 each simulation $N_{\rm FRB}=5000$ FRBs are generated by using the flat
 $w$CDM model with the preset cosmological parameters $\Omega_m=0.3153$ and
 $w=-0.95$. The marginalized $1\sigma$ constraints are on the cosmological
 parameter $\Omega_m$.}
 \end{figure}
 \end{center}

%======================================================================

 \vspace{-22.4mm}  % used here just for a comfortable typesetting for Figs. 5 and 6

 \be{eq17}
 {\rm DM_{HG}}={\rm DM_{HG\cmn loc}}/(1+z)\,,
 \ee
 where $\rm DM_{HG\cmn loc}$ is the local DM of FRB host galaxy. Following
 e.g.~\cite{Yang:2017bls,Qiang:2019zrs,Qiang:2020vta}, we reasonably assume
 that $\rm DM_{HG\cmn loc}$ is independent of redshift $z$.

We briefly describe the steps to generate the simulated FRBs
 with ``\,known\,'' redshifts. At first, we assign a random redshift $z_i$
 to the $i$-th simulated FRB from a given redshift distribution (one of
 the nine mentioned in Sec.~\ref{sec2}). In this step, the distributions
 Pzconst and PzSFH should use a given cosmology characterized
 by $E(z)=H(z)/H_0$ (which will be specified in Sec.~\ref{sec4}) to
 calculate the comoving and luminosity distances, while the
 other distributions should not. As mentioned above, both the intergalactic
 hydrogen and helium are fully ionized at $z\leq 3$, and hence we choose to
 generate the FRB redshifts in the range of $0\leq z_i\leq 3$
 (namely $z_{\rm max}=3$). The second step is to randomly
 assign ${\rm DM}_{{\rm IGM}\cmn i}$ and its uncertainty
 $\sigma_{{\rm IGM}\cmn i}=\sigma_{\rm IGM}(z_i)$ to this simulated FRB
 from a Gaussian distribution,
 \be{eq18}
 {\rm DM}_{{\rm IGM}\cmn i}={\cal N}(\langle
 {\rm DM_{IGM}}\rangle (z_i)\cmn\sigma_{\rm IGM}(z_i))\,.
 \ee
 Here, $\langle{\rm DM_{IGM}}\rangle (z_i)$ in Eq.~(\ref{eq14})
 is calculated by using a given cosmology characterized by $E(z)$ (which
 will be specified in Sec.~\ref{sec4}), and $\sigma_{\rm IGM}(z_i)$ is
 calculated by using Eq.~(\ref{eq16}). The third step is to
 assign ${\rm DM}_{{\rm HG}\cmn i}={\rm DM}_{{\rm HG\cmn loc}
 \cmn i}/(1+z_i)$ and its uncertainty $\sigma_{{\rm HG}\cmn i}=
 \sigma_{{\rm HG\cmn loc}\cmn i}/(1+z_i)$ to this simulated FRB, according
 to Eq.~(\ref{eq17}) and following e.g.~\cite{Yang:2016zbm,Gao:2014iva,
 Yang:2017bls,Zhou:2014yta,Qiang:2019zrs,Qiang:2020vta,Li:2019klc}. Here,
 ${\rm DM}_{{\rm HG\cmn loc}\cmn i}$ can be randomly assigned from a
 Gaussian distribution with the mean $\langle{\rm DM_{HG\cmn loc}}\rangle$
 and a fluctuation $\sigma_{\rm HG\cmn loc}$ \cite{Yang:2016zbm,Gao:2014iva,
 Yang:2017bls,Zhou:2014yta,Qiang:2019zrs,Qiang:2020vta,Li:2019klc}, namely
 \be{eq19}
 {\rm DM}_{{\rm HG\cmn loc}\cmn i}={\cal N}(\langle{\rm DM_{HG
 \cmn loc}}\rangle\cmn\sigma_{\rm HG\cmn loc})\,,
 \ee
 while $\sigma_{{\rm HG\cmn loc}\cmn i}=\sigma_{\rm HG\cmn loc}$. In the
 literature, ${\rm DM_{HG\cmn loc}}=50\;\dmunit$ is frequently used (see
 e.g.~\cite{Shannon:2018,Prochaska:2019,Hashimoto:2019aqu}). On the other
 hand, it was argued in~\cite{Zhang:2020mgq} that the median of
 ${\rm DM_{HG\cmn loc}}$ is about $30\sim 70\;\dmunit$, while
 the uncertainty $20\;\dmunit$ was frequently used in the literature
 (e.g.~\cite{Yang:2016zbm,Qiang:2019zrs,Qiang:2020vta}). So, we adopt the
 fiducial values $\langle{\rm DM_{HG\cmn loc}}\rangle=50\;\dmunit$ and
 $\sigma_{\rm HG\cmn loc}=20\;\dmunit$ in this work. Finally, the simulated
 $\rm DM_E$ data and its uncertainty for the $i$-th simulated FRB are
 given by
 \be{eq20}
 {\rm DM}_{{\rm E}\cmn i}={\rm DM}_{{\rm IGM}\cmn i}+{\rm DM}_{{\rm HG}
 \cmn i}\,,\quad\quad {\rm and}\quad\quad
 \sigma_{{\rm E}\cmn i}=(\sigma_{{\rm IGM}\cmn i}^2+\sigma_{{\rm HG}
 \cmn i}^2)^{1/2}\,.
 \ee
 One can repeat the above steps for $N_{\rm FRB}$ times
 to generate $N_{\rm FRB}$ simulated FRBs.

The lower-limit estimates for the number of FRB events are a few thousands
 per sky per day~\cite{Keane:2018jqo,Bhandari:2017qrj}. Even conservatively,
 the all-sky burst rate floor derived from the pre-commissioning of
 CHIME/FRB is $3\times 10^2$ events per day \cite{Amiri:2019qbv}. Several
 projects designed to detect and localize FRBs with arcsecond accuracy in
 real time are under construction or in commission, for example
 DSA-10~\cite{DSA-10}, DSA-2000~\cite{DSA-2000},
 MeerKAT~\cite{Jankowski:2020lqz}, UTMOST-2D~\cite{UTMOST}, and
 LOFAR~\cite{vanHaarlem:2013dsa}. It is reasonable to expect that numerous
 FRBs with identified redshifts will become available in the future.
 Therefore, $N_{\rm FRB}$ can be fairly large, for example
 ${\cal O}(10^3)$ or even more.

%============================= Fig. 7 =================================

 \begin{center}
 \begin{figure}[tb]
 \centering
 \vspace{-6mm} \hspace{-6mm} % used here just for a comfortable typesetting
 \includegraphics[width=0.82\textwidth]{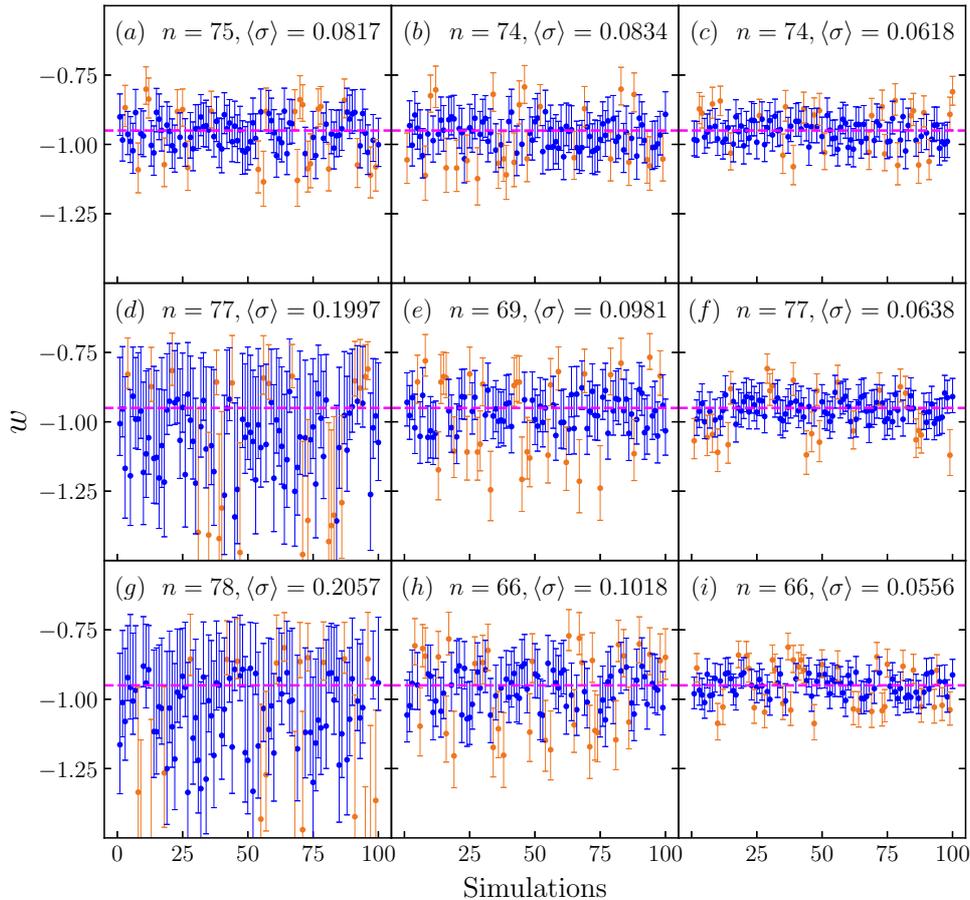}
 \caption{\label{fig7} The same as in Fig.~\ref{fig6}, but the marginalized
 $1\sigma$ constraints are on the cosmological parameter $w$.}
 \end{figure}
 \end{center}

%======================================================================

%============================= Fig. 8 =================================

 \begin{center}
 \begin{figure}[tb]
 \centering
 \vspace{-10mm} \hspace{-6mm} % used here just for a comfortable typesetting
 \includegraphics[width=0.82\textwidth]{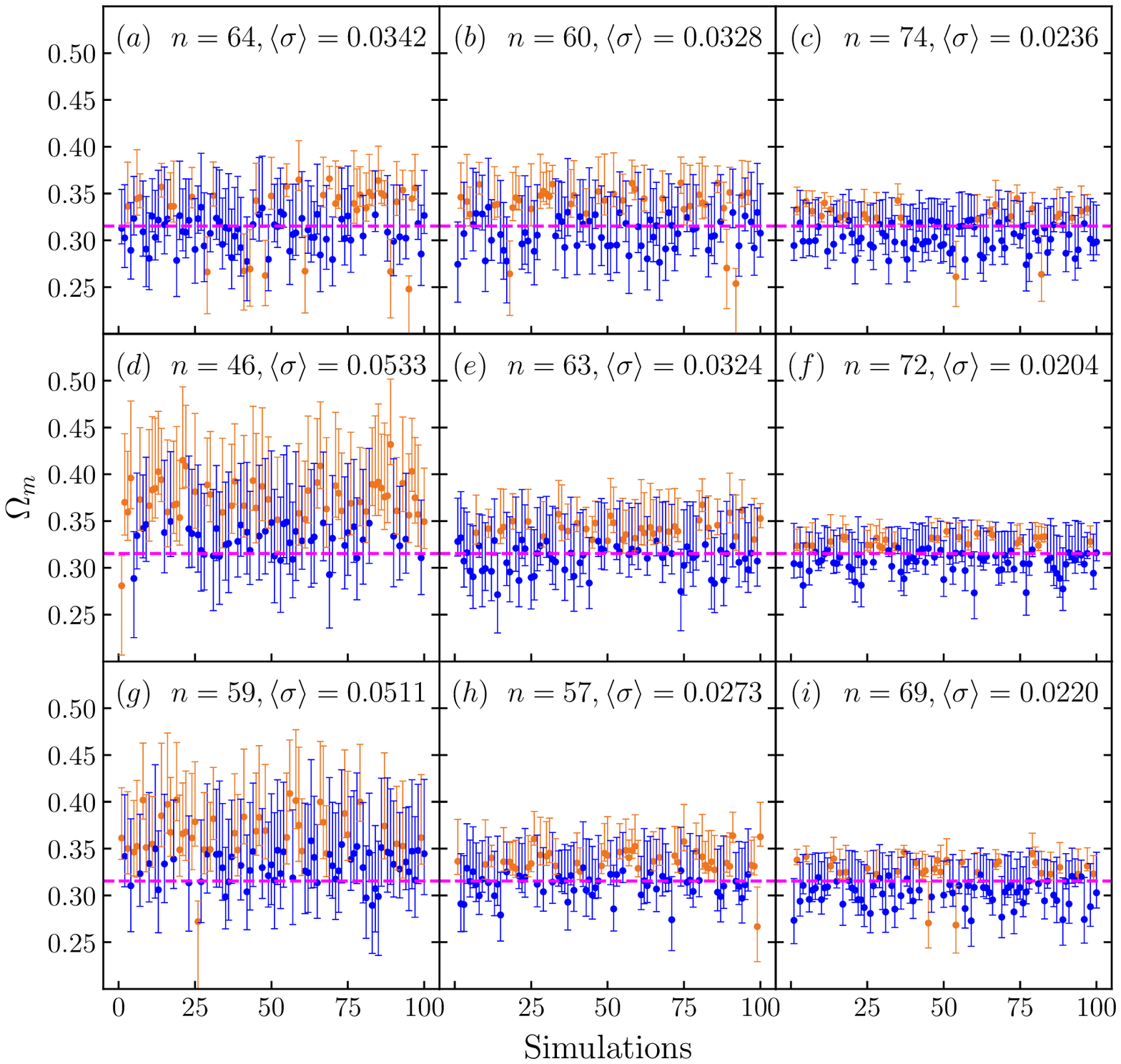}
 \caption{\label{fig8} The same as in Fig.~\ref{fig3}, but in
 each simulation $N_{\rm FRB}=5000$ FRBs are generated by using the flat
 CPL model with the preset cosmological parameters $\Omega_m=0.3153$,
 $w_0=-0.95$ and $w_a=-0.3$. The marginalized $1\sigma$ constraints are
 on the cosmological parameter $\Omega_m$.}
 \end{figure}
 \end{center}

%======================================================================

%============================= Fig. 9 =================================

 \begin{center}
 \begin{figure}[tb]
 \centering
 \vspace{-9mm} \hspace{-6mm} % used here just for a comfortable typesetting
 \includegraphics[width=0.82\textwidth]{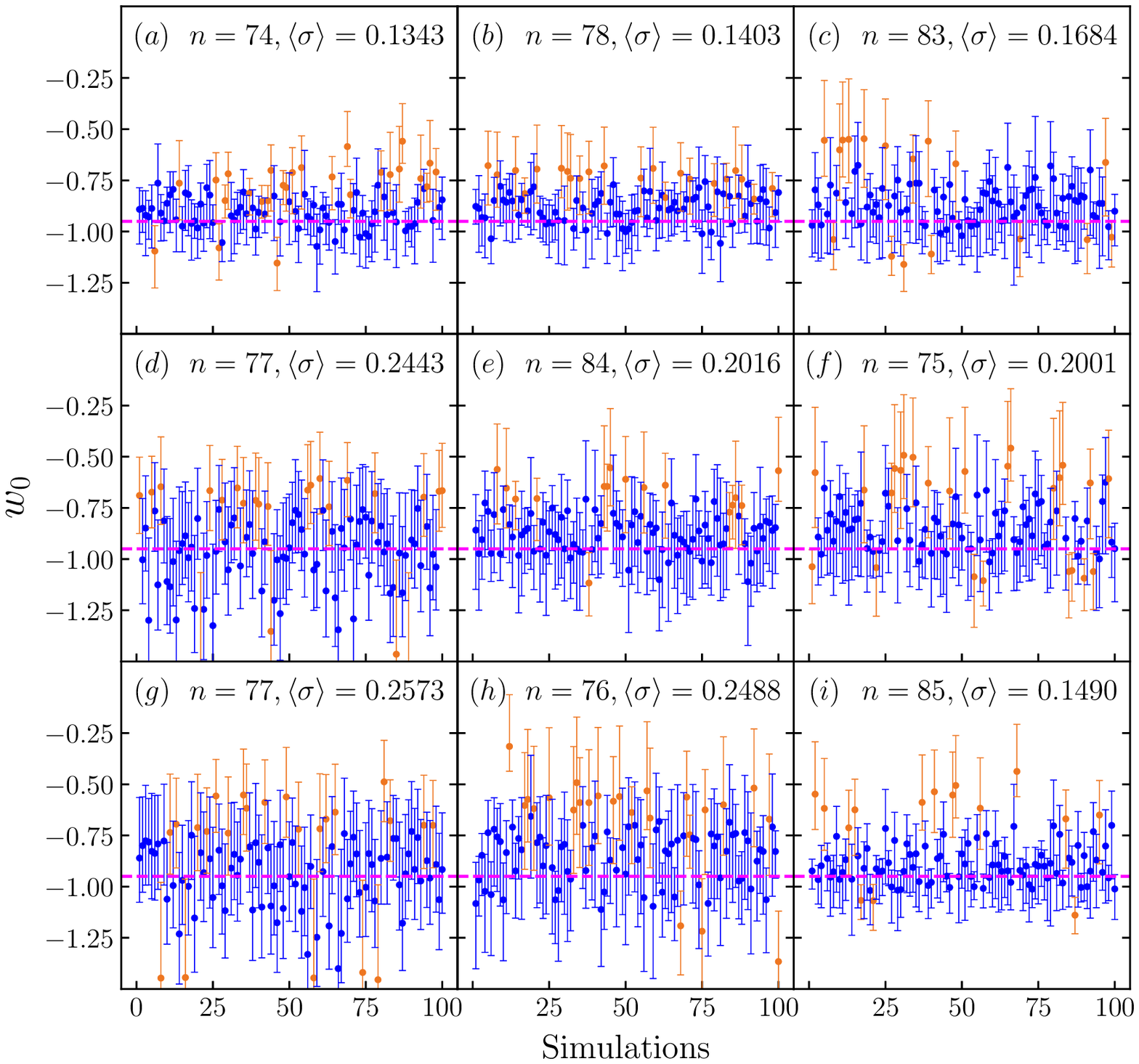}
 \caption{\label{fig9} The same as in Fig.~\ref{fig8}, but the marginalized
 $1\sigma$ constraints are on the cosmological parameter $w_0$.}
 \end{figure}
 \end{center}

%======================================================================

%============================= Fig. 10 =================================

 \begin{center}
 \begin{figure}[tb]
 \centering
 \vspace{-9mm} \hspace{-6mm} % used here just for a comfortable typesetting
 \includegraphics[width=0.82\textwidth]{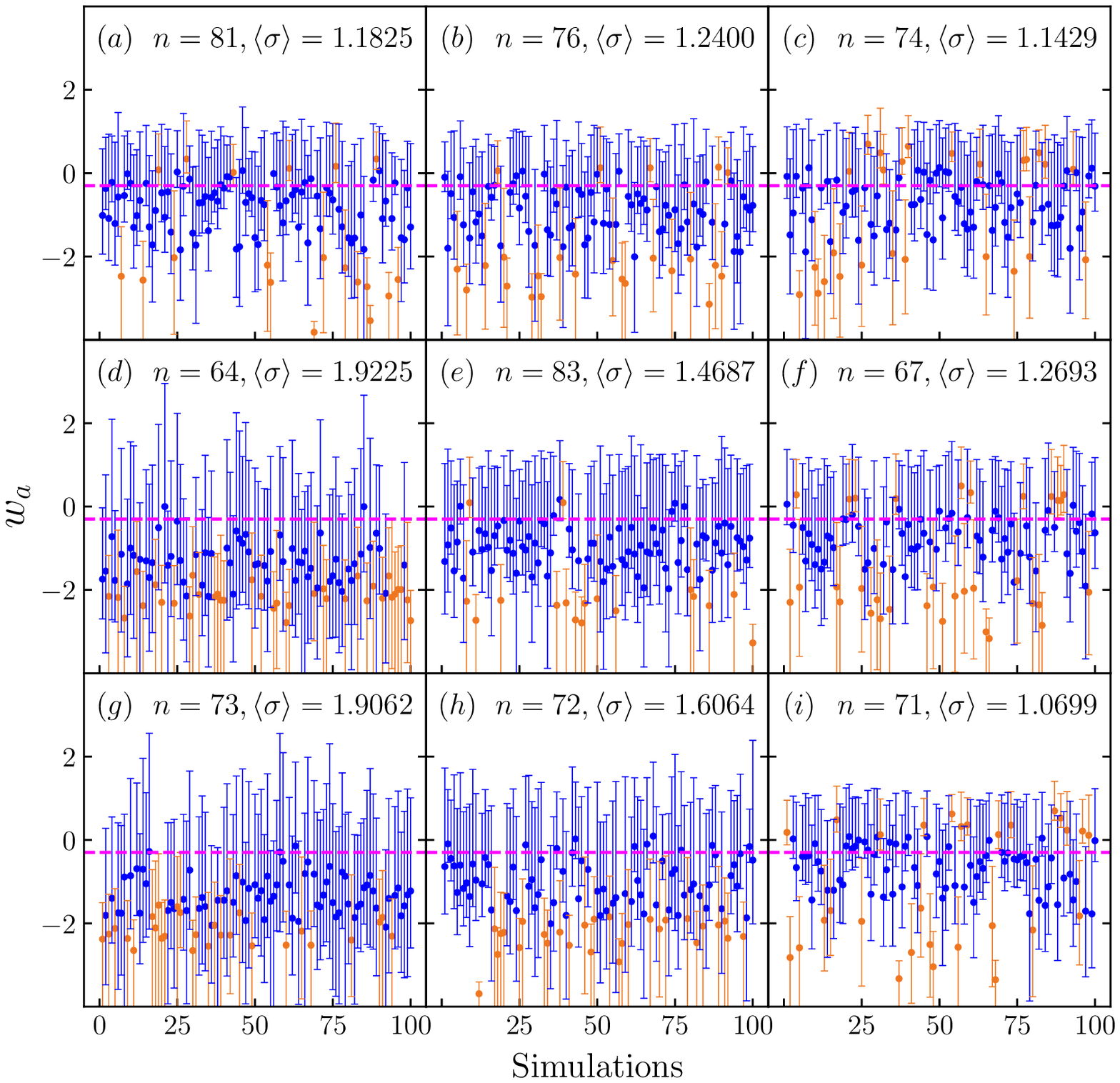}
 \caption{\label{fig10} The same as in Fig.~\ref{fig8}, but the marginalized
 $1\sigma$ constraints are on the cosmological parameter $w_a$.}
 \end{figure}
 \end{center}

%======================================================================

 \vspace{-37mm}  % used here just for a comfortable typesetting for Figs. 7 -- 10

%============================= section 4 ===================================

\section{Cosmological constraints from the simulated FRBs}\label{sec4}

Now, we consider the constraints on various cosmological models from
 the simulated FRBs. For a specified cosmological model, its dimensionless
 Hubble parameter $E(z)=H(z)/H_0$ is given. So, one can calculate the
 theoretical extragalactic DM of an FRB by using
 \be{eq21}
 {\rm DM_E^{th}}(z)=\langle{\rm DM_{IGM}}\rangle(z)+\langle{\rm DM_{HG\cmn
 loc}}\rangle /(1+z)\,,
 \ee
 where $\langle{\rm DM_{IGM}}\rangle(z)$ is given by Eq.~(\ref{eq14}),
 and the universal constant $\langle{\rm DM_{HG\cmn loc}}\rangle$ is a
 model parameter for HG. The model parameters can
 be constrained by performing a $\chi^2$ analysis, while
 \be{eq22}
 \chi^2=\sum_i\frac{\left({\rm DM}_{{\rm E}
 \cmn i}-{\rm DM_E^{th}}(z_i)\right)^2}{\sigma_{{\rm E}\cmn i}^2}\,.
 \ee
 In this work, we use the Markov Chain Monte Carlo (MCMC) code
 {\bf CosmoMC}~\cite{Lewis:2002ah} to this end. Since we are mainly
 interested in the effect of redshift distributions on cosmological
 constraints, to save the length of paper, we do not present
 the constraints on HG parameter $\langle{\rm DM_{HG\cmn loc}}\rangle$
 in the following, although they are also available in fact.

At first, we consider the simplest cosmological model, namely the flat
 $\Lambda$CDM model. In this case, the dimensionless Hubble parameter
 is given by (e.g.~\cite{Wei:2010wu,Liu:2014vda})
 \be{eq23}
 E(z)=(\Omega_m (1+z)^3+(1-\Omega_m))^{1/2}\,,
 \ee
 where $\Omega_m$ is the only free cosmological parameter. We simulate
 $N_{\rm FRB}$ FRBs with the preset cosmological parameter $\Omega_m=0.3153$
 taken from Planck 2018 results~\cite{Aghanim:2018eyx}. Then, we constrain
 the flat $\Lambda$CDM model with these simulated FRBs. To avoid the
 statistical noise due to random fluctuations, one should repeat the
 constraints for a large number of simulations. However, it is fairly
 expensive to consider too many simulations since they consume a large
 amount of computation power and time. As a balance, we choose to consider
 100 simulations, which is enough in fact.

In Fig.~\ref{fig3}, the marginalized $1\sigma$ constraints on
 the cosmological parameter $\Omega_m$ of the flat $\Lambda$CDM model for
 100 simulations are presented. In each simulation, $N_{\rm FRB}=1000$ FRBs
 are generated. It is easy to see from Fig.~\ref{fig3} that the preset
 parameter $\Omega_m=0.3153$ can be found within $1\sigma$ region in most of
 the 100 simulations ($64\sim77\%$), for all cases of the 9 redshift
 distributions introduced in Sec.~\ref{sec2}. This implies that the
 cosmological constraints from simulated FRBs are fairly reliable and
 robust. However, the uncertainties of the constraints are different. Using
 the naked eye, we find from Fig.~\ref{fig3} that the error bars of right
 panels $(c)$, $(f)$, $(i)$ are shortest, the ones of bottom-left panels
 $(d)$, $(g)$ are longest, and the ones of other four panels $(a)$, $(b)$,
 $(e)$, $(h)$ are moderate. Quantitatively, $\langle\sigma\rangle$ in each
 panel gives the mean of the uncertainties of 100 constraints. Using
 $\langle\sigma\rangle$ in Fig.~\ref{fig3}, we confirm that the
 cosmological constraints from FRBs simulated with the redshift
 distributions $(c)$~PzGRB, $(f)$~PzSFR, $(i)$~Uniform are tightest, the
 ones with the redshift distributions $(d)$~Pzconst(0.5), $(g)$~PzSFH(0.5)
 are loosest, and the ones with the redshift distributions $(a)$~Burr,
 $(b)$~Burr12, $(e)$~Pzconst(1.0), $(h)$~PzSFH(1.0) are moderate. Clearly,
 they are separated into three distinct groups.

In Figs.~\ref{fig4} and \ref{fig5}, the number of simulated FRBs increases
 to $N_{\rm FRB}=5000$ and 10000, respectively. Clearly, the cosmological
 constraints become tighter when the number of simulated FRBs increases,
 for all cases of the 9 redshift distributions. However, the insight about
 the constraining ability keeps unchanged. The FRBs simulated with the
 redshift distributions PzGRB/PzSFR/Uniform, Pzconst(0.5)/PzSFH(0.5),
 Burr/Burr12/Pzconst(1.0)/PzSFH(1.0) have strong, weak,
 moderate constraining abilities, respectively. These three groups of
 redshift distributions lead to different cosmological constraining
 abilities from the simulated FRBs. Using FRB simulations with different
 redshift distributions, one will make optimistic, pessimistic,
 or moderate predictions about the future of the FRB cosmology.

Let us turn to other cosmological models to see whether or not
 the above insight changes. The second is the flat $w$CDM
 model, in which the dimensionless Hubble parameter is
 given by (e.g.~\cite{Wei:2010wu,Liu:2014vda})
 \be{eq24}
 E(z)=\left[\,\Omega_m (1+z)^3+(1-\Omega_m)(1+z)^{3(1+w)\,}\right]^{1/2}\,,
 \ee
 where $\Omega_m$ and $w$ are free cosmological parameters. We simulate
 $N_{\rm FRB}$ FRBs with the preset cosmological parameter $\Omega_m=0.3153$
 and $w=-0.95$. Then, we constrain the flat $w$CDM model with these
 simulated FRBs. In Figs.~\ref{fig6} and \ref{fig7}, the marginalized
 $1\sigma$ constraints on the cosmological parameters $\Omega_m$ and
 $w$ are presented, respectively. Since the insight about the constraining
 ability keeps unchanged when the number of simulated FRBs varies, we
 only consider the case of $N_{\rm FRB}=5000$ for the flat $w$CDM model.
 Once again, it is easy to see from Figs.~\ref{fig6} and \ref{fig7} that
 both the preset parameters $\Omega_m=0.3153$ and $w=-0.95$ can be found
 within $1\sigma$ region in most of the 100 simulations, for all cases of
 the 9 redshift distributions introduced in Sec.~\ref{sec2}. This implies
 that the cosmological constraints from simulated FRBs are fairly reliable
 and robust. On the other hand, since there are two free cosmological
 parameters $\Omega_m$ and $w$ in the flat $w$CDM model while there is only
 one cosmological parameter $\Omega_m$ in the flat $\Lambda$CDM model,
 the constraints on $\Omega_m$ in the flat $w$CDM model (Fig.~\ref{fig6})
 are looser than the ones in the $\Lambda$CDM model (Fig.~\ref{fig4}), as
 expected. From Figs.~\ref{fig6} and \ref{fig7}, one can find that the
 cosmological constraints on both $\Omega_m$ and $w$ from FRBs
 simulated with the redshift distributions PzGRB/PzSFR/Uniform,
 Pzconst(0.5)/PzSFH(0.5), and Burr/Burr12/Pzconst(1.0)/PzSFH(1.0) are
 tightest, loosest, and moderate, respectively. These three groups of
 redshift distributions lead to different cosmological constraining
 abilities from the simulated FRBs. This insight still holds in the case
 of flat $w$CDM model.

Finally, we consider the flat Chevallier-Polarski-Linder (CPL)
 model~\cite{Chevallier:2000qy,Linder:2002et}, in which the
 equation-of-state parameter (EoS) of dark energy is parameterized as
 \be{eq25}
 w = w_0 + w_a \left(1-a\right)= w_0 + w_a\,\frac{z}{1+z}\,,
 \ee
 where $w_0$ and $w_a$ are constants. As is well known, the corresponding
 $E(z)$ is given by (e.g.~\cite{Wei:2010wu,Liu:2014vda})
 \be{eq26}
 E(z)=\left[\,\Omega_{m}(1+z)^3+\left(1-\Omega_{m}\right)
 (1+z)^{3(1+w_0+w_a)}\exp\left(-\frac{3w_a z}{1+z}\right)\right]^{1/2}\,.
 \ee
 We simulate $N_{\rm FRB}$ FRBs with the preset parameters
 $\Omega_m=0.3153$, $w_0=-0.95$ and $w_a=-0.3$. Then, we constrain the
 flat CPL model with these simulated FRBs. Similarly, we only consider
 the case of $N_{\rm FRB}=5000$ for the flat CPL model. In
 Figs.~\ref{fig8}--\ref{fig10}, the marginalized $1\sigma$ constraints on
 the cosmological parameters $\Omega_m$, $w_0$ and $w_a$ are presented,
 respectively. Again, we find from Figs.~\ref{fig8}--\ref{fig10} that
 the preset cosmological parameters $\Omega_m=0.3153$, $w_0=-0.95$ and
 $w_a=-0.3$ can be found within $1\sigma$ region in most of the
 100 simulations, for almost all cases of the 9 redshift distributions
 introduced in Sec.~\ref{sec2}. There are 3 free cosmological parameters
 in this model, and hence the cosmological constraints will become worse
 than the flat $\Lambda$CDM and $w$CDM models which have 1 and 2 free
 cosmological parameters, respectively. This can be verified by
 comparing Figs.~\ref{fig8}--\ref{fig10} with Figs.~\ref{fig4}, \ref{fig6}
 and \ref{fig7}.

Let us look at the uncertainties of cosmological constraints. In the
 case of $\Omega_m$ (Fig.~\ref{fig8}), the same insight keeps unchanged
 as in the flat $\Lambda$CDM and $w$CDM models, namely the cosmological
 constraints on $\Omega_m$ from FRBs simulated with the
 redshift distributions PzGRB/PzSFR/Uniform, Pzconst(0.5)/PzSFH(0.5),
 and Burr/Burr12/Pzconst(1.0)/PzSFH(1.0) are tightest, loosest,
 and moderate, respectively.

However, it is slightly changed in the cases of $w_0$ and $w_a$. The
 tightest, loosest, and moderate groups are changed to the distributions
 Burr/Burr12/PzGRB/Uniform, Pzconst(0.5)/PzSFH(0.5)/PzSFH(1.0), and
 Pzconst(1.0)/PzSFR in the case of $w_0$ (Fig.~\ref{fig9}), respectively.
 On the other hand, the tightest, loosest, and moderate groups are changed
 to the distributions Burr/Burr12/PzGRB/PzSFR/Uniform,
 Pzconst(0.5)/PzSFH(0.5), and Pzconst(1.0)/PzSFH(1.0) in the case of $w_a$
 (Fig.~\ref{fig10}), respectively. This is mainly due to
 the correlation between the cosmological parameters $w_0$ and $w_a$.

Nevertheless, it is still unchanged that different redshift distributions
 lead to different cosmological constraining abilities from the simulated
 FRBs. Thus, if one uses the unsuitable redshift distributions to simulate
 FRBs, rather than the actual one of FRBs (which is still unknown to date),
 overoptimistic or overpessimistic predictions about the future of the
 FRB cosmology might be made.

%============================= section 5 ===================================

\section{Concluding remarks}\label{sec5}

Nowadays, FRBs have been a promising probe for astronomy and cosmology.
 However, it is not easy to identify the redshifts of FRBs to date.
 Thus, no sufficient actual FRBs with identified redshifts can be used
 to study cosmology currently. In the past years, one has to use the
 simulated FRBs with ``\,known\,'' redshifts instead. To simulate an
 FRB, one should randomly assign a redshift to it from a given redshift
 distribution. But the actual redshift distribution of FRBs is still
 unknown so far. Therefore, various redshift distributions have been
 assumed in the literature, while some of them are motivated by the star
 formation history/rate or compact binary mergers and so on, some of them
 are borrowed from other objects such as gramma-ray bursts (GRBs), some of
 them come from the observed FRBs, and some of them are not well motivated
 at all. In the present work, we study the effect of various redshift
 distributions on cosmological constraints, and we do not care whether these
 redshift distributions are well motivated or where they come from. Our goal
 is just to see how they affect the cosmological constraints, while they are
 treated equally in this work, no matter whether they are the intrinsic ones
 or the observed ones. We find that different redshift distributions lead to
 different cosmological constraining abilities from the simulated FRBs. This
 result emphasizes the importance to find the actual resdshift distribution
 of FRBs, and reminds us of the possible bias in the FRB simulations due
 to the redshift distributions.

In fact, one should also include the contribution from Galactic halos
 into $\rm DM_{obs}$ in Eq.~(\ref{eq2}). Although it is poorly known,
 ${\rm DM_{MW,\,halo}}\approx 50\sim 80\,\dmunit$ was suggested
 in e.g.~\cite{Prochaska:2019mn}. Actually, $\rm DM_{MW,\,halo}$ can be
 absorbed into $\rm DM_{MW}$ and then Eq.~(\ref{eq2}) takes the same form
 (but with a different meaning for $\rm DM_{MW}$), because we only use
 the extragalactic DM (namely $\rm DM_E\equiv DM_{IGM}+DM_{HG}$ as defined
 in Eq.~(\ref{eq3})) to constrain the cosmological models.

We stress that there are other types of redshift distributions
 for FRBs in the literature. We do not try to consider all redshift
 distributions for FRBs in a limited work. However, it is expected that
 our main conclusion will not change for the other redshift
 distributions unused in this work.

As mentioned above, the 9 redshift distributions can be separated into
 three distinct groups, namely PzGRB/PzSFR/Uniform, Pzconst(0.5)/PzSFH(0.5),
 and Burr/Burr12/Pzconst(1.0)/PzSFH(1.0), which lead to strong, weak, and
 moderate constraining abilities, respectively. In fact, we can find
 some clues from the left panel of Fig.~\ref{fig2}. The normalized redshift
 distributions PzGRB/PzSFR/Uniform are commonly ``\,short and wide\,'', and
 hence the simulated FRBs span almost the whole redshift range from 0 to 3.
 On the contrary, the normalized redshift distributions
 Pzconst(0.5)/PzSFH(0.5) are commonly ``\,tall and thin\,'' with a sharp
 peak nearby the low redshift $0.5$, and hence the simulated
 FRBs concentrate in a narrow redshift range around the redshift $0.5$
 (in fact it is rare to have redshifts $>1$). On the other hand, the
 normalized redshift distributions Pzconst(1.0)/PzSFH(1.0) are moderate,
 and hence the simulated FRBs span a fairly wide redshift range from 0 to
 $\sim$2.2. Although the normalized redshift distributions Burr/Burr12
 tilt to low redshifts $<1$, they have not so small probability to generate
 redshifts in the range from 1 to 2. In this sense, Burr/Burr12 are
 similar to Pzconst(1.0)/PzSFH(1.0), so that they are also in the moderate
 group. These are clues found from the left panel of Fig.~\ref{fig2}.
 Naively, we try to understand them as follows. For FRBs at high redshifts,
 $\rm DM_E$ is dominated by the contribution from IGM, namely
 $\rm DM_{IGM}$, which carries the key information about IGM and the cosmic
 expansion history. In this case, the contribution from host galaxy,
 ${\rm DM_{HG}=DM_{HG\cmn loc}}/(1+z)$, becomes relatively small. So, it is
 expected that the constraints on the cosmic expansion history accordingly
 become tight. Thus, the redshift distributions having larger probability to
 generate redshifts $>2$ or $>1$ lead to stronger cosmological constraining
 abilities (smaller uncertainties) from the simulated FRBs.

In this work, we have proposed two new redshift distributions, namely
 Burr and Burr12, from the observed FRBs to date. It should be emphasized
 that the observed redshift distributions is a convolution of the intrinsic
 redshift distribution (which is unknown so far), the FRB luminosity/energy
 distribution (which is also unknown), and the sensitivity of telescopes (we
 thank the referee for pointing out this issue). On the other hand, many
 selection effects exist in the observed FRBs from different telescopes.
 Thus, Burr and Burr12 redshift distributions cannot be taken seriously, and
 cannot be mixed with the intrinsic redshift distribution (which is unknown
 so far). However, even if we remove Burr, Burr12 and Uniform from the 9
 redshift distributions considered in this work, the other 6 redshift
 distributions still lead to the same conclusion. They are still separated
 into three distinct groups, namely PzGRB/PzSFR, Pzconst(0.5)/PzSFH(0.5),
 and Pzconst(1.0)/PzSFH(1.0), without taking Burr, Burr12 and Uniform into
 account. Although all of the other 6 redshift distributions come from some
 intrinsic redshift distribution models such as star formation history, they
 still lead to distinct cosmological constraints, as shown in the present
 work. Burr and Burr12 are not in the center of this work, and they also
 cannot change the central conclusion. In fact, some authors used various
 redshift distributions (such as PzGRB, PzSFR, Pzconst and PzSFH) to
 simulate FRBs and accordingly considered the constraints on cosmological
 models in the literature. They made some claims with one of these assumed
 redshift distributions (such as PzGRB, PzSFR, Pzconst and PzSFH). Our
 point is that such kind of claims in the literature considerably depend
 on the assumed redshift distributions. We follow their steps but with
 various assumed redshift distributions, and find that different redshift
 distributions lead to different claims. So, such kind of works in the
 literature are not robust, since there is bias in their claims. This is
 our key point. To this problem, the key is to find the actual intrinsic
 resdshift distribution of FRBs, which is still unknown to date. We hope
 that it will be available in the near future. Of course, one can still
 make some helpful efforts in this direction. For example, in
 e.g.~\cite{Zhang:2020ass,James:2021oep}, several intrinsic redshift
 distribution models (tracking the star formation history/rate or compact
 binary mergers and so on) were tested with the observational data, and
 found that they are consistent with these data currently. We consider
 this is an important topic in the field of FRBs, especially
 for the FRB cosmology.

%============================= acknowledgements ==================================

\section*{ACKNOWLEDGEMENTS}

We thank the anonymous referee for quite useful comments and suggestions,
 which helped us to improve this work. We are grateful to Hua-Kai~Deng,
 Shu-Ling~Li, Zhong-Xi~Yu, Han-Yue~Guo and Jing-Yi~Jia for kind help and
 useful discussions. This work was supported in part by NSFC under Grants
 No.~11975046 and No.~11575022.

\renewcommand{\baselinestretch}{1.0}

%============================= references =================================

\end{document}